\documentclass[aps,showpacs,epsfig,twocolumn]{revtex4}
\usepackage{epsfig}

\newcommand{\be}{\begin{equation}}
\newcommand{\ee}{\end{equation}}
\newcommand{\bea}{\begin{eqnarray}}
\newcommand{\eea}{\end{eqnarray}}
\begin{document}

\title{\bf\Large {Three-Body Collective Excitations in the Superconducting Phase of $MgB_2$}}

\author{Naoum Karchev\cite{byline}}

\affiliation{Department of Physics, University of Sofia, 1164 Sofia, Bulgaria}

\begin{abstract}
It is shown that coherent behavior of Cooper's pairs and electrons in the two-band superconductors $MgB_2$ is possible as a result of a strong two-phonon-electron coupling.    
The spin-$\frac 12$ triples with zero angular momentum are made up of three spin-$\frac 12$ fermions with charge $\emph{e}$. They are gapped Fermi excitations with gap induced by the gaps of the single fermions. Their contribution to the thermodynamics of the $MgB_2$ superconductors is considered. 

\end{abstract}


\maketitle 

{\bf I INTRODUCTION}
\ \\

The three-body bound states are everywhere in the physics, but the study of their role in collective condensed matter behavior is still limited. Phase transitions driven by an instability in a three-fermion channel have been explored. 
The anomalous three-body scattering amplitude is used as an ansatz to develop a mechanism of odd-frequency superconductivity \cite{trip02}. Unlike a BCS theory, the authors involve a cooperative pairing of electrons and spins. The three-body bound state formation leads to gap which is an odd function of frequency. The application to heavy fermion superconductors is discussed. 

In the present paper the  interest in this topic is inspired  by 
the anomalous superconducting properties of $MgB_2$ \cite{trip1}. The superconductivity in magnesium diboride is $s$ wave, mediated by electron-phonon coupling. It differs from ordinary metallic superconductors in several ways. Scanning tunneling microscopy \cite{trip2} and point contact studies \cite{trip3} revealed double-peaked spectra at low temperature that were interpreted as evidence for two gap superconductivity. \emph{Ab initio} calculations suggest that multiple gaps are a consequence of the coupling of distinct electronic bands \cite{trip4,trip4a}. $MgB_2$ has strongly anisotropic Fermi surface of four separate sheets that are grouped into two-dimensional $\sigma$ bands and three-dimensional $\pi$ bands. The different energy gaps 
are associated with the $\pi$ and $\sigma$ sheets. 
The key quantity in phonon-mediated superconductivity is the Eliashberg function $\alpha^2(\omega)F(\omega)$,  where $F(\omega)$ is phonon density of states and $\alpha(\omega)$ is the electron-phonon coupling averaged over directions in $k$ space.
The electron tunneling spectroscopy is used \cite{trip03} to determine the three distinct Eliashberg functions $(\alpha^2F)_{\pi-\pi}$,  $(\alpha^2F)_{\sigma-\sigma}$, and $(\alpha^2F)_{\pi-\sigma}$.

The double-gap structure is used to explain some of the unusual physical properties of $MgB_2$, such as 
the rapid rise of the specific heat coefficient $C/T$ \cite{trip4b}, tunneling \cite{trip3} and upper critical field anisotropy \cite{trip4c}. Quite different methods of theoretical investigation, including weak-coupling two-band BCS theory \cite{trip4d}, Eliashberg strong-coupling formalism \cite{trip4a}, and  strong-coupling density-functional technique with explicit account for the Coulomb repulsion \cite{trip4e}, lead to astonishingly identical curves for specific heat, as a function of temperature. The calculations reproduce the different slopes, above $0.5Tc$ and below $0.25Tc$, which apparently result from the existence of two gaps, but can not explain the shoulder between them \cite{trip4b,trip4f}. Unexpected are the effects of $Mg$ substitution by $Al$ on the specific heat. The changes are in rather poor agreement with those predicted by taking into account changes in the electronic and photonic structure only \cite{trip4g}.      

A large $B$ isotope effect is another argument in favor of phonon-mediated pairing \cite{trip5,trip5a}. 
The isotope coefficient $\alpha$ is defined by the relation $T_{c} \propto M^{-\alpha}$, where $M$ is the mass of the element. In BCS theory $\alpha=0.5$, and for metals like $Hg$, $Pb$ and $Zn$ the coefficient is found experimentally to be close to $0.5$. The isotope coefficient for $MgB_2$ is $\alpha\approx0.32$.
The density-functional calculations of the phonon spectrum and electron-phonon coupling in $MgB_2$ predict that in this compound, phonon modes of Boron oscillations may have relatively high frequencies, and that nonlinear coupling via two-phonon exchange is comparable to or even larger than the linear coupling \cite{trip6,trip4}.
Both effects may contribute to the anomalous  isotope effect coefficient \cite{trip5a}, and to the significant increasing of the critical temperature $T_c=39K$. It is thought that $T_c$ of $MgB_2$ probably represents the upper
limit of the phonon mediated superconductivity. 

Motivated by the theoretical and experimental findings I consider theory of two-band superconductors  \cite{trip7,trip8} with two-phonon electron interaction. The main goal is to study the formation of coherent behavior of Cooper's pairs and electrons in the two-band superconductors $MgB_2$, as a result of a strong two-phonon-electron coupling. It is shown that spin-$\frac 12$ triples with zero angular momentum, made up of three 
spin-$\frac 12$ fermions with charge $\emph{e}$, are possible. They are gapped Fermi excitations with gap induced by the gaps of the single fermions. Effectively one can represent them as gapped fermions and to write an effective action. This enable to calculate the contribution of the triples to the thermodynamics of the $MgB_2$ superconductors.
To reproduce the shoulder in the specific heat as a function of temperature, one has to choose the gaps of the triples larger than the gaps of the incipient electrons. 

The paper is organized as follows. In Sec. II the formation of triples and their contribution to the specific heat is explored. A summary in Sec. IV concludes this paper. The tunneling experiments are debated as a most direct way to observe the triples experimentally.   

\ \\

{\bf II TRIPLES IN TWO-BAND SUPERCONDUCTORS WITH NONLINEAR ELECTRON-PHONON COUPLING}
\ \\

\begin{figure}[h]  
\vspace{0.1cm}  
\epsfxsize=4.5cm  
\hspace*{0.2cm}  
\epsfbox{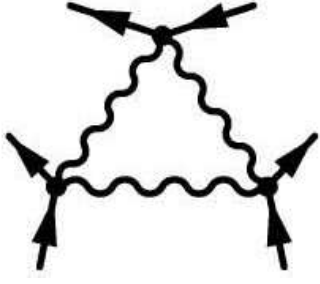}   
\caption{Triangle diagram with three phonon lines (undulating lines) which results from the two-phonon exchange}  
\label{fig1}  
\end{figure}

I consider theory of two-band superconductors  \cite{trip7,trip8} with two-phonon electron interaction.  An important consequence of this interaction is the effective six-fermion interaction. One can obtain it from the triangle diagram (Fig\ref{fig1}) with three phonon lines (undulating lines). 
There are two fermion species in the theory, therefore the low frequency and low momenta limit of the diagram leads to an effective local six-fermion term. The effective Hamiltonian $H_{f^6}$ of the six-fermion interaction has the form 
\bea\label{trip1}
& H_{f^6} = \\ 
&-\sum\limits_{\ell\sigma}\lambda_{\ell}\int d^3x
c^+_{\ell\uparrow}(\textbf{x})c^+_{\ell\downarrow}(\textbf{x})c^+_{-\ell\sigma}(\textbf{x})
c_{-\ell\sigma}(\textbf{x})c_{\ell\downarrow}(\textbf{x})c_{\ell\uparrow}(\textbf{x})
\nonumber
\eea
where $c^+_{\ell\sigma}(\textbf{x})$ and $c_{\ell\sigma}(\textbf{x})$ are creation 
and annihilation operators for fermions, with orbital index $\ell$ 
($\ell=1,2,\,-\ell=2,1$) and spin projection $\sigma$ ($\sigma=\uparrow,\downarrow$) \cite{trip9}.

The BCS Hamiltonian of the theory of two-band superconductivity is \cite{trip7,trip8}
\bea\label{trip2}
& H_{BCS}=-\sum\limits_{\ell}\,g_{\ell}\,\int d^3x c^+_{\ell\uparrow}(\textbf{x})c^+_{\ell\downarrow}(\textbf{x})
c_{\ell\downarrow}(\textbf{x})c_{\ell\uparrow}(\textbf{x}) \\
& -g_3 \int d^3x \sum\limits_{\ell}c^+_{\ell\uparrow}(\textbf{x})c^+_{\ell\downarrow}(\textbf{x})
c_{-\ell\downarrow}(\textbf{x})c_{-\ell\uparrow}(\textbf{x}) \nonumber
\eea

The partition function can be written as a path integral over the Grassmann  
functions of the Matsubara time $\tau$\,\,\,\,$c^+_{\ell\sigma}(\tau,\textbf{x})$ and $c_{\ell\sigma}(\tau,\textbf{x})$
\be\label{trip3}
{\cal Z}(\beta)\,=\,\int\,D\mu\left(c^+\,c\right) e^{-S}.
\ee
The action is given by the expressions  
\bea 
& S= S_0+\int\limits^{\beta}_0 d\tau H_{\text{int}}(\tau) \\
& S_0= \int\limits^{\beta}_0 d\tau \int d^3x \sum\limits_{\ell}
c^+_{\ell\sigma}(\tau,\textbf{x})\left[\partial_{\tau}+\epsilon_{\ell}(\nabla)\right]
c_{\ell\sigma}(\tau,\textbf{x})
\eea
where $\beta$ is the inverse temperature and $\epsilon_{\ell}(\nabla)$ is the dispersion of band $\ell$ fermion. 
The Hamiltonian $H_{\text{int}}(\tau)$ is a sum of the BCS Hamiltonian (\ref{trip2}) 
and six-fermion Hamiltonian (\ref{trip1}). It is obtained 
from Eqs.(\ref{trip1}) and (\ref{trip2}) replacing the operators with Grassmann functions. 

Let us introduce two spin-$\frac 12$ fermi collective fields (\textbf{triples}) 
$\zeta_{\ell\sigma}(\tau,\textbf{x})\,(\zeta^+_{\ell\sigma}(\tau,\textbf{x}))$ by means of the Hubbard-Stratanovich transformation of the six-fermion term (\ref{trip1}) 
\bea\label{trip4}
e^{-H_{f^6}} = 
\int
D\mu(\zeta^+\zeta)  
\exp\{-\int d^4x\sum\limits_{\ell} \lambda_{\ell}\left[
\zeta^+_{\ell\sigma}(x)\zeta_{\ell\sigma}(x)+\right.\nonumber\\\\
\left.  c^+_{\ell\uparrow}(x)c^+_{\ell\downarrow}(x)c^+_{-\ell\sigma}(x)
\zeta_{\ell\sigma}(x)+\zeta^+_{\ell\sigma}(x)
c_{-\ell\sigma}(x)c_{\ell\downarrow}(x)c_{\ell\uparrow}(x)\right]\}
\nonumber
\eea
where $x=(\tau,\textbf{x})$ and $\int\limits^{\beta}_0 d\tau\int d^3x=\int d^4x$.

The effective action of the triples is defined by the equality 
\bea\label{trip5}
e^{-S_{\text{\emph{eff}}}(\zeta^+,\zeta)}=
\exp\left\{-\int d^4x\sum\limits_{\ell\sigma}\lambda_{\ell}
\zeta^+_{\ell\sigma}(x)\zeta_{\ell\sigma}(x)\right\} \\
<\exp\{-\int d^4x\sum\limits_{\ell\sigma}\lambda_{\ell}\left[c^+_{\ell\uparrow}(x)c^+_{\ell\downarrow}(x)c^+_{-\ell\sigma}(x)
\zeta_{\ell\sigma}(x)+ \right. \nonumber \\
\left. \zeta^+_{\ell\sigma}(x)
c_{-\ell\sigma}(x)c_{\ell\downarrow}(x)c_{\ell\uparrow}(x)\right]\}>
\nonumber
\eea
with
\be\label{trip6}
<Q>=\int D\mu(c^+,c)\,Q\,e^{-S_0-H_{\text{BCS}}}
\ee
The quadratic part of the effective action $S_{\text{\emph{eff}}}(\zeta^+,\zeta)$ 
has the form  
\bea\label{trip7}
S_{eff}=\int d^4x d^4y\left[\zeta^+_{\ell\sigma}(x)\Pi^{\ell\ell'}_{\sigma\sigma'}(x-y)\zeta_{\ell'\sigma'}(y)+
\right. \\
\left.\zeta^+_{\ell\sigma}(x)\Sigma^{\ell\ell'}_{\sigma\sigma'}(x-y)\zeta^+_{\ell'\sigma'}(y)+
\zeta_{\ell\sigma}(x)\overline{\Sigma}^{\ell\ell'}_{\sigma\sigma'}(x-y)\zeta_{\ell'\sigma'}(y)\right].
\nonumber
\eea
where
\bea\label{trip8}
\Pi^{\ell\ell'}_{\sigma\sigma'}(x-y)= 
\delta_{\ell,\ell'}\delta_{\sigma\sigma'}\lambda_{\ell}\delta^4(x-y)- \nonumber\\
\lambda_{\ell}\lambda_{\ell'} <c_{-\ell\sigma}(x)c_{\ell\downarrow}(x)c_{\ell\uparrow}(x)
c^+_{\ell'\uparrow}(y)c^+_{\ell'\downarrow}(y)c^+_{-\ell'\sigma'}(y)>  \\
\Sigma^{\ell\ell'}_{\sigma\sigma'}(x-y)= \nonumber \\
\frac {\lambda_{\ell}\lambda_{\ell'}}{2} <c_{-\ell\sigma}(x)c_{\ell\downarrow}(x)c_{\ell\uparrow}(x)
c_{-\ell'\sigma'}(y)c_{\ell'\downarrow}(y)c_{\ell'\uparrow}(y)> \\
\overline{\Sigma}^{\ell\ell'}_{\sigma\sigma'}(x-y)= \nonumber \\
\frac {\lambda_{\ell}\lambda_{\ell'}}{2} <c^+_{\ell\uparrow}(x)c^+_{\ell\downarrow}(x)c^+_{-\ell\sigma}(x)
c^+_{\ell'\uparrow}(y)c^+_{\ell'\downarrow}(y)c^+_{-\ell'\sigma'}(y)> 
\eea
In theory with Hamiltonian $H_{BCS}$ (\ref{trip2}) the off-diagonal elements are zero: $\Pi^{\ell\ell'}_{\sigma\sigma'}=0$,
$\Sigma^{\ell\ell'}_{\sigma\sigma'}=0$ and $\overline{\Sigma}^{\ell\ell'}_{\sigma\sigma'}=0$ if $\ell\neq\ell'$.
The diagonal functions are calculated in leading order represented by the diagrams (Fig2). In the case of 
$\Pi^{\ell\ell}_{\sigma\sigma'}$ (Fig2a), two lines of the diagram (solid lines) correspond to the normal Green function of fermions with one and just the same band-index, while the third line (dashed line) corresponds to the normal Green function of fermion with different band-index. The diagrams for $\Sigma^{\ell\ell}_{\sigma\sigma'}$ and  $\overline{\Sigma}^{\ell\ell}_{\sigma\sigma'}$ (Fig2b) have two lines (solid lines) corresponding to the anomalous Green functions of fermions with equal band-index and one line (dashed line) which corresponds to the anomalous Green function of fermion with different band-index. 
\begin{figure}[h]  
\vspace{0.55cm}  
\epsfxsize=5.50cm  
\hspace*{0.2cm}  
\epsfbox{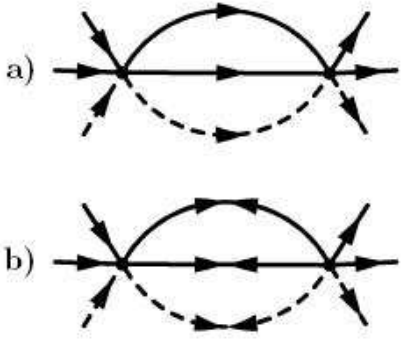}   
\caption{Two loop diagrams which represent leading order of: a) $\Pi^{\ell\ell}_{\sigma\sigma'}$, 
b) $\Sigma^{\ell\ell}_{\sigma\sigma'}$ and  
$\overline{\Sigma}^{\ell\ell}_{\sigma\sigma'}$}  
\vspace{-0.5cm} 
\label{fig}  
\end{figure}  

The system of gaps' equations in the theory of two-band superconductivity has the form \cite{trip7,trip8}
\bea\label{tripseq}
\Delta_1=\frac {g_1}{2}\int d^3p\frac {\tanh(\frac {\beta}{2} E_{1\textbf{p}})}{E_{1\textbf{p}}}\Delta_1+
\frac {g_3}{2}\int d^3p\frac {\tanh(\frac {\beta}{2}E_{2\textbf{p}})}{E_{2\textbf{p}}}\Delta_2 
\nonumber \\ \\
\Delta_2=\frac {g_3}{2}\int d^3p\frac {\tanh(\frac {\beta}{2}E_{1\textbf{p}})}{E_{1\textbf{p}}}\Delta_1+
\frac {g_2}{2}\int d^3p\frac {\tanh(\frac {\beta}{2}E_{2\textbf{p}}}{E_{2\textbf{p}})}\Delta_2 
\nonumber
\eea
with $E_{\ell\textbf{p}}=\sqrt {\varepsilon^2_{\ell\textbf{p}}+|\Delta^2_{\ell}|}$, where $\epsilon_{\ell\textbf{p}}$ is $\ell$-band fermion dispersion. The system of gaps' equations shows that the gaps $\Delta_{\ell}$ can be chosen real. Then 
the normal and anomalous Green functions in two-band theory, calculated in the standard way, have the form 
\cite{trip10}
\bea\label{trip9}
\textit{S}^{\ell}_{\uparrow\uparrow}(\omega,\textbf{p})=S^{\ell}_{\downarrow\downarrow}(\omega,\textbf{p})=
-\frac {i\omega+\varepsilon_{\ell\textbf{p}}}{\omega^2+\varepsilon^2_{\ell\textbf{p}}+\Delta^2_{\ell}} \\
\textit{F}(\omega,\textbf{p})=\textit{F}^+(\omega,\textbf{p})=\frac {\Delta_{\ell}}{\omega^2+\varepsilon^2_{\ell\textbf{p}}+\Delta^2_{\ell}}
\label{trip10}
\eea
I calculate the diagrams in low-frequency limit 
\be\label{trip10a}
\Pi^{\ell\ell}_{\sigma\sigma'}(\omega,\textbf{p})=\delta_{\sigma\sigma'}\left[-i\omega Z^{-1}_{\ell}(\textbf{p})+\hat{\varepsilon}_{\ell}(\textbf{p})\right]
\ee
\be\label{trip10b} 
\Sigma^{\ell\ell}_{\downarrow\uparrow}(0,\textbf{p})=
\overline{\Sigma}^{\ell\ell}_{\uparrow\downarrow}(0,\textbf{p})=\Sigma_{\ell}(\textbf{p}).
\ee
The result is 
\newpage
\bea\label{trip11}
Z^{-1}_{\ell}(\textbf{p})= \nonumber \\ 
\int\prod\limits_{i=1}^3\frac {d^3p_i}{(2\pi)^3}
\frac {(2\pi)^3\delta^3(\textbf{p}_1+\textbf{p}_2+\textbf{p}_3-\textbf{p})}
{4E_{-\ell\textbf{p}_1}E_{\ell\textbf{p}_2}E_{\ell\textbf{p}_3}
(E_{-\ell\textbf{p}_1}+E_{\ell\textbf{p}_2}+E_{\ell\textbf{p}_3})} \nonumber \\
\left[E_{-\ell\textbf{p}_1}E_{\ell\textbf{p}_2}E_{\ell\textbf{p}_3}+
E_{-\ell\textbf{p}_1}\varepsilon_{\ell\textbf{p}_2}\varepsilon_{\ell\textbf{p}_3}+\right. \nonumber\\
\left. \varepsilon_{-\ell\textbf{p}_1}E_{\ell\textbf{p}_2}\epsilon_{\ell\textbf{p}_3}+
\epsilon_{-\ell\textbf{p}_1}\epsilon_{\ell\textbf{p}_2}E_{\ell\textbf{p}_3}\right]
\nonumber \\
\eea
\bea\label{trip12}
\hat{\varepsilon}_{\ell}(\textbf{p})=\frac {1}{\lambda_{\ell}}+ \nonumber \\ 
\int\prod\limits_{i=1}^3\frac {d^3p_i}{(2\pi)^3}
\frac {(2\pi)^3\delta^3(\textbf{p}_1+\textbf{p}_2+\textbf{p}_3-\textbf{p})}
{4E_{-\ell\textbf{p}_1}E_{\ell\textbf{p}_2}E_{\ell\textbf{p}_3}
(E_{-\ell\textbf{p}_1}+E_{\ell\textbf{p}_2}+E_{\ell\textbf{p}_3})} \nonumber \\
\left[\varepsilon_{-\ell\textbf{p}_1}\varepsilon_{\ell\textbf{p}_2}\varepsilon_{\ell\textbf{p}_3}+
\varepsilon_{-\ell\textbf{p}_1}E_{\ell\textbf{p}_2}E_{\ell\textbf{p}_3}+\right.\nonumber \\
\left. E_{-\ell\textbf{p}_1}\varepsilon_{\ell\textbf{p}_2}E_{\ell\textbf{p}_3}+
E_{-\ell\textbf{p}_1}E_{\ell\textbf{p}_2}\varepsilon_{\ell\textbf{p}_3}\right]
\nonumber \\
\eea
\bea\label{trip13}
\Sigma_{\ell}(\textbf{p})= \nonumber \\ 
\frac 12\int\prod\limits_{i=1}^3
\frac {d^3p_i}{(2\pi)^3}
\frac {\Delta_{\ell}^2\Delta_{-\ell}(2\pi)^3\delta^3(\textbf{p}_1+\textbf{p}_2+\textbf{p}_3-\textbf{p})}
{4E_{-\ell\textbf{p}_1}E_{\ell\textbf{p}_2}E_{\ell\textbf{p}_3}
(E_{-\ell\textbf{p}_1}+E_{\ell\textbf{p}_2}+E_{\ell\textbf{p}_3})} \nonumber \\
\eea

The equations 
\be\label{trip13a}
\hat{\varepsilon}_{\ell}(\tilde {p}_{f\ell})=0
\ee 
define the Fermi surface of the triples.
Using the approximate expressions for $Z_{\ell}(\textbf{p})$ and $\Sigma_{\ell}(\textbf{p})$
\be\label{trip13b}
Z_{\ell}(\textbf{p})\simeq Z_{\ell}(p_{f\ell})=Z_{\ell},\qquad
\Sigma_{\ell}(\textbf{p})\simeq\Sigma_{\ell}(p_{f\ell}), 
\ee
and re-scaling the triples' fields
\bea\label{trip13c}
& & Z^{-\frac 12}\zeta^+_{\ell\sigma}(\omega,\textbf{p})\rightarrow\zeta^+_{\ell\sigma}(\omega,\textbf{p}),\\
& & Z^{-\frac 12}\zeta_{\ell\sigma}(\omega,\textbf{p})\rightarrow\zeta_{\ell\sigma}(\omega,\textbf{p}) 
\eea
one obtains the effective action of the triples 
\bea\label{trip14}
S_{eff}=\int\limits^{\beta}_0 d\tau\sum\limits_{p}\left\{\zeta^+_{\ell\sigma}(\tau,\textbf{p})\left(\partial_{\tau}
+\tilde{\varepsilon}_{\ell}(\textbf{p})\right)\zeta_{\ell\sigma}(\tau,\textbf{p})+\right. \\
\left.\varrho_{\ell}\left[\zeta^+_{\ell\uparrow}(\tau,\textbf{p})\zeta^+_{\ell\downarrow}(\tau,\textbf{-p})+
\zeta_{\ell\downarrow}(\tau,\textbf{-p})\zeta_{\ell\uparrow}(\tau,\textbf{p})\right] \right\}
\nonumber
\eea
with $\tilde{\varepsilon}_{\ell}(\textbf{p})=Z_{\ell}\hat{\varepsilon}_{\ell}(\textbf{p})$ and
$\varrho_{\ell}=Z_{\ell}\Sigma_{\ell}(\textbf{p}_{f\ell})$. Near the Fermi surface
$\tilde{\varepsilon}_{\ell}(\textbf{p})\simeq\frac {\tilde {p}_{f\ell}}{\tilde {m}_{\ell}}(p-p_{f\ell})$, where $\tilde {m}_{\ell}$ are triples' masses. 

The effective action (\ref{trip14}) shows that triples are spin-$\frac 12$ fermi excitations with gap $\varrho_{\ell}$ induced by the gaps of the single fermion excitations
(\ref{trip13}). Next one diagonalizes the effective Hamiltonian using a Bogoliubov transformation, and rewrites it in terms of Bogoliubov excitations with dispersion $\tilde {E}_{\ell}(p)=\sqrt {\tilde\varepsilon^2_{\ell}(p)+\varrho^2_{\ell}}$. This enable to calculate the contribution of triples to the thermodynamics of superconductors. In particular the low temperature behavior of the heat capacity is   
\be\label{trip15}
C_{s}=\sum\limits_{\ell}\left[\frac {m_{\ell}p_{f\ell}}{\pi^2}
\sqrt {\frac {2\pi\Delta^5_{\ell}}{T^3}}e^{-\frac {\Delta_{\ell}}{T}} +
\frac {\tilde {m}_{\ell}\tilde {p}_{f\ell}}{\pi^2}
\sqrt {\frac {2\pi\rho^5_{\ell}}{T^3}}e^{-\frac {\rho_{\ell}}{T}} 
\right]
\ee 
where the first terms come from the single fermion contribution, while the other terms are triples' contribution. 
Recent experiments, including photoemission \cite{trip11} and tunneling experiments \cite{trip2,trip12}, show that the ratio of the single fermions' gaps is $2.6\leq\Delta_2/\Delta_1\leq3.5$. The low temperature behavior of the heat capacity coefficient
$\frac {\pi^2}{m_1 p_{f1}\sqrt {2\pi}}\frac {C}{T}=\tilde {\gamma}$ is depicted as a function of $\frac {T}{\Delta_1}$ in Fig.3 for
$\frac {\Delta_2}{\Delta_1}=3, \frac {\varrho_1}{\Delta_1}=4, \frac {\varrho_2}{\Delta_1}=5$ and
$m_1 p_{f1}\simeq m_2 p_{f2}\simeq \tilde {m}_1\tilde {p}_{f1}\simeq \tilde {m}_2\tilde {p}_{f2}$.
The curve is in a good agreement with the experimental one for $MgB_2$ \cite{trip4b,trip4f,trip4g}. 
\begin{figure}[h]  
\vspace{-2cm}  
\epsfxsize=8.5cm  
\hspace*{-1.cm}  
\epsfbox{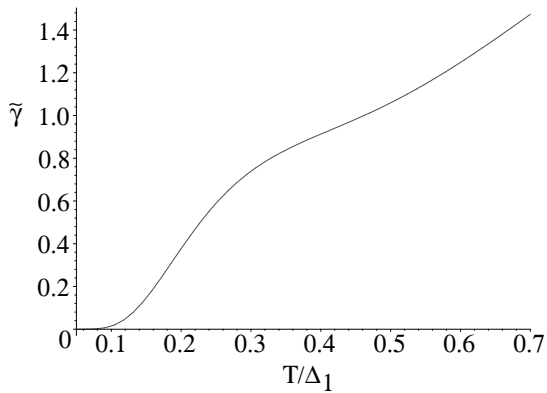}   
\vspace{-3.5cm}
\caption{The heat capacity coefficient
$\frac {\pi^2}{m_1 p_{f1}\sqrt {2\pi}}\frac {C}{T}=\tilde {\gamma}$ as a function of $\frac {T}{\Delta_1}$
 for $\frac {\Delta_2}{\Delta_1}=3, \frac {\varrho_1}{\Delta_1}=4, \frac {\varrho_2}{\Delta_1}=5$ and
$m_1 p_{f1}\simeq m_2 p_{f2}\simeq \tilde {m}_1\tilde {p}_{f1}\simeq \tilde {m}_2\tilde {p}_{f2}$.}  
\label{fig3}  
\vspace{-0.5cm}  
\end{figure} 
 
\ \\

{\bf III SUMMARY}
\ \\

In summary, new type of excitations, triples- made up of three spin-$\frac 12$ fermions, are predicted in theory of two-band superconductivity with non-linear two-phonon-fermion interaction. They can be thought of as a bound-state of s-type Cooper pair of fermions from one of the bands and fermion from the other one with zero angular momentum. The triples are gapped excitations with gap induced by the single fermion gaps. It is important to underline that the triples result from the nonlinear two-phonon-electron interaction. One can consider excitations made up of more than three spin-$\frac 12$ fermions while studying diagrams with more than three vertexes. Due to the Pauli principle these excitations have non-zero angular momentum.  The specific form of the fermion interactions determine the symmetry of these excitations. They in turn result from the non-linear two-phonon-fermion interaction. Therefore the knowledge of the non-linear phonon interaction is crucial for the development of a theory of excitations made up of more than three fermions. The same is true for triples made up of fermions from one band. They have non-zero angular momentum too, and the symmetry of the triples results from the non-linear two-phonon-fermion interaction. 

The six-fermion interaction (\ref{trip1}) can be alternatively obtained from the linear phonon-electron coupling. 
The leading order diagram is depicted in Fig4.
\begin{figure}[h]  
\vspace{0.02cm}  
\epsfxsize=4.5cm  
\hspace*{0.2cm}  
\epsfbox{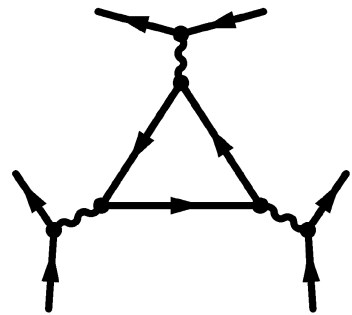}   
\caption{Six-fermion interaction obtained from the linear phonon-electron coupling}  
\label{fig4}  
\end{figure}   
Since my investigation is based on the assumption that nonlinear coupling via two-phonon exchange is comparable to or even larger than the linear coupling \cite{trip6,trip4}, the contribution of the linear phonon-electron coupling (Fig4) is small perturbation to the contribution of the nonlinear phonon-electron interaction and one can drop it.

It is easy to see from the formula for the heat capacity (\ref{trip15}) that the contribution of the triples is decisive to explain the shoulder like part of the curve. 

The most promising way to observe the triples in $MgB_2$ is by tunneling experiments. 
It is evident that the contribution of the triples to the tunneling current is with much smaller weight than those of the single electrons. Therefore, we can observe the triples only at very low temperature. 
If one considers the differential conductance of tunnel junction
with counter-electrodes in superconducting state, for example $In$, or $Pb$ as in \cite{trip03},  two new peaks should emerge when the temperature decreases.
Tunneling experiments achieved bellow $1K$ can answer the question about the existence or nonexistence of the triples.

\end{document}